# Design of efficient single stage chirped pulse difference frequency generation at 7 μm driven by a dual wavelength Ti:sapphire laser

**Christian Erny**[1,2,*] **and Christoph P. Hauri**[1,2,3]

[1]*Paul Scherrer Institute, SwissFEL, 5232 Villigen PSI, Switzerland*
[2]*Ecole Polytechnique Federale de Lausanne, 1015 Lausanne, Switzerland*
[3]*christoph.hauri@psi.ch*
[*]*Christian.erny@psi.ch*

**Abstract:** We present a design for a high-energy single stage mid-IR difference frequency generation adapted to a two-color Ti:sapphire amplifier system. The optimized mixing process is based on chirped pulse difference frequency generation (CP-DFG), allowing for a higher conversion efficiency, larger bandwidth and reduced two photon absorption losses. The numerical start-to-end simulations include stretching, chirped pulse difference frequency generation and pulse compression. Realistic design parameters for commercially available non linear crystals (GaSe, $AgGaS_2$, $LiInSe_2$, $LiGaSe_2$) are considered. Compared to conventional un-chirped DFG directly pumped by Ti:sapphire technology we report a threefold increase of the quantum efficiency. Our CP-DFG scheme provides up to 340 μJ pulse energy directly at 7.2 μm when pumped with 3 mJ and supports a bandwidth of up to 350 nm. The resulting 240 fs mid-IR pulses are inherently phase stable.

## 1. Introduction

Intense laser pulses tunable in the mid infrared wavelength range (3 to 20 μm) are interesting for numerous applications, ranging from investigations of the fingerprint spectral region and semiconductors [1,2] to the scaling of high order harmonic generation towards the water window and beyond [3–5]. A prominent approach for accessing this wavelength range is based on a multi-stage white-light seeded optical parametric amplifier (OPA) system, driven by a femtosecond Ti:sapphire laser system. While such BBO-based OPA stages typically provide tunable output between 1.3 μm and 2.6 μm by signal and idler, the longer wavelength range (3-20 μm) is usually accessed by an additional difference frequency generation (DFG) stage [6–8] of signal and idler. For such tunable systems the typical conversion efficiency between the near-IR and the IR is rather small (≈1%), the corresponding quantum efficiency (QE) little (≈10%), the energy stability rather poor (5-10% rms after DFG) and the pulses not intrinsically CEP stable.

In this paper we present a novel approach of a single stage, high-energy difference frequency generation which is based on chirped phase matching [9] of two intense laser pulses with different colors from a common Ti:sapphire amplifier. The scheme offers several advantages. It provides an increased energy stability, a simple experimental setup, directly carrier envelope phase (CEP) stable [10,11] mid-IR output pulses, as well as an increased phase matching bandwidth and higher conversion efficiency. In particular, the chirped input beams significantly reduce the impact of two-photon absorption (TPA) due to lower input intensity compared to the un-chirped case.

Our investigation is motivated by recent progress in Ti:sapphire amplifier [12] which is now capable to deliver synchronously a pair of intense pulses which are easily tunable in their central wavelengths. Thanks to the advancement of intra-cavity acousto-optic pulse shaping [11] in the regenerative amplifier, tunable two and more color operation up to 20 mJ has been demonstrated in our group [14]. The large wavelength separation between two colors (up to 90 nm) makes this source well suited for direct DFG with only one nonlinear mixing stage covering potentially the spectral range between 7-30 μm. Such a single DFG stage approach is favorable in view of stability, conversion efficiency and reduced complexity. As example, our two-color amplifier system [14] provides a typical energy stability of 0.54% rms and similar stable DFG output is thus expected since the energy fluctuations at the output of the DFG are directly proportional to the energy fluctuations of the driving laser field.

In the past, several DFG approaches based on AGS and GS have been presented. Throughout those efforts un-chirped (transform-limited (TL)) femtosecond pulses were used. Due to the high intensities and the subsequently strong TPA only a low DFG efficiency could be achieved which limited the generated output power significantly. Even though the applied laser systems provided pulse energies at 800 nm comparable to the laser parameter assumed here, the generated output energy between 7 - 10 μm was not exceeding 8 μJ with AGS [15,16], corresponding to a pump to idler QE of 3.7%, and not higher than 0.6 μJ in GS (QE=3.4%) between 10-20 μm [17]. Most known nonlinear materials show relatively high TPA at around 800 nm when irradiated with intense femtosecond pulses. This limited the applicable energy substantially in the past.

To our best knowledge, DFG with chirped input pulses from a Ti:sapphire laser has not been investigated up to present. DFG with chirped pulses, however, offers the potential for significantly reduced TPA effect and offers thus higher output energies. We performed start to end simulations for chirped pulse difference frequency generation (CP-DFG) under realistic conditions, based on the above mentioned two-color Ti:sapphire laser system. This includes chirping, nonlinear wave mixing with two photon absorption and subsequent pulse compression. The principal goal was to find pump power, chirp, crystal thickness, beam size, and compression scheme for broadband and efficient DFG. We demonstrate that by CP-DFG a quantum efficiency of up to 60% can be achieved while maintaining a large bandwidth. This exceeds largely the quantum efficiency of previous experimental implementations and of the conventional multistage OPA approach. We show that the generated pulses can be compressed close to the transform limit by direct bulk material compressing.

We have structured this paper the following way. First we discuss the limitations of DFG with transform limited pulses (TL-DFG) and demonstrate performance improvements achieved by CP-DFG (section 2), followed by a detailed description of our multidimensional optimization process (section 3) and a brief discussion on temporal pulse compression after CP-DFG (section 4).

All simulations in this paper have been performed with the nonlinear propagation code by Arisholm [18]. The same code has been successfully used to model optical parametric chirped pulse amplifiers (OPCPA, [19]) at 800 nm [20,21] and 3.5 μm [22,23] and chirped sum frequency generation [24]. It numerically solves the equations for second-order nonlinear frequency mixing for a full three-dimensional beam in an arbitrary birefringent crystal and takes into account the effects of depletion, diffraction, walk-off, and TPA.

## 2. DFG with transform-limited and chirped pulses

As driving source for DFG a multi-millijoule, dual-wavelength Ti:sapphire laser system is considered, operating at 760 nm (pump, TL 53 fs) and 850 nm (signal, TL 66 fs), with a bandwidth of 16 nm for both pulses. The given maximum wavelength separation allows generating mid-IR radiation at 7.2 μm (idler). The case study in this paper addresses in detail this particular mixing process but in principle, by reducing the wavelength separation between pump and signal, the idler could be tuned easily to longer wavelength. Our study is based on the commercially available nonlinear materials $AgGaS_2$ (AGS), GaSe (GS), $LiInSe_2$ (LISe), and $LiGaSe_2$ (LGSe). While AGS and GS are well established, LISe and LGSe have only become available commercially recently.

As a benchmark, we first calculated conventional DFG with TL pump and TL signal for different DFG crystals in dependence of the total input intensity. The calculations shown in Fig. 1. provide the DFG output energy at a wavelength of 7.2 µm as function of applied intensity. The generated idler energy is for a $1/e^2$ beam radius for pump and signal of 5 mm. It turns out that even at high intensities of 50 GW/cm$^2$ and with a total input energy of 4 mJ not more than 40 µJ output energy is achieved (Fig. 1a.). These numerical results are in well agreement with the previously reported experimental results, corroborating that the main limiting factor in TL-DFG for a larger conversion into the idler is the high loss through TPA, as shown in Fig. 1b. The loss by TPA increases rapidly even at low intensities (i.e. a few GW/cm$^2$) and surpasses 80% for intensities above 40 GW/cm$^2$ (LISe, AGS and GS). The large and detrimental impact of TPA in the mixing process can only be controlled by reducing the input intensity. While a reduction in intensity by reducing the pump energy goes along with a reduction of the DFG output we propose to chirp the two input pulses while keeping the pump pulse energy high. This chirped-pulse DFG (CP-DFG) requires, however, careful investigation and control of the wavelength-dependent mixing process to achieve efficient conversion and broadband phase-matching.

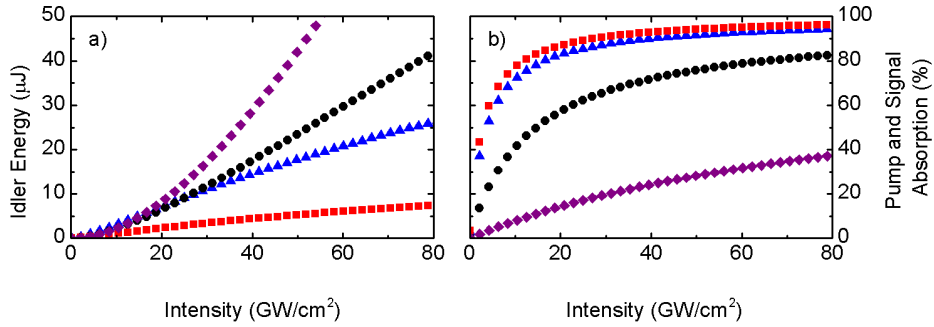

Fig. 1. Conventional DFG with transform-limited pulses. Achievable output power (a) and pump and signal absorption (b) for DFG with compressed pulses in 10 mm diameter 2 mm long AGS, (red squares), 1 mm long GS (blue triangles), 3 mm long LISe (black dots), and 2 mm long LGSe (purple diamonds).

The benefit of CP-DFG over TL-DFG is illustrated in Fig. 2. by presenting best-effort achievements for both. Since the intensity is the driving force behind the two competing processes of nonlinear wave mixing and TPA, we optimized chirp and total intensity as well as crystal length to enhance the broadband DFG output. The results indicate clearly that the DFG output energy can be increased by more than an order of magnitude by inducing a specific chirp to pump and signal. Additionally CP-DFG enhances the spectral full width at half maximum (FWHM) bandwidth (Fig. 2c.) compared to TL-DFG (Fig. 2d.).

Fig. 2. illustrates furthermore that CP-DFG operates at significantly lower input intensity for efficient DFG output compared to TL-DFG. For each intensity value plotted in Fig. 2a. an optimized set of pump chirp, pump/signal chirp ratio, total input energy and pump/signal intensity ratio is used whose optimization is discussed in detail in section 3.

To get a realistic estimation for the overall performance of CP-DFG we performed in addition time-consuming 3D simulations at the critical intensity, i.e. the intensity where the maximum tolerable TPA occurs (see section 3). The 3D simulations is fed with the optimized pump chirp, chirp ratio and pump/signal intensity ratio provided by 1D simulations and found by the corresponding parameter scans shown in section 3. The 3D results for nonlinear mixing are shown in Table 1 and give a comprehensive overview on optimum applied intensity, chirps, input energies and the resulting output energy and bandwidth. These results take into account the commercially available crystal size, being 10 mm in diameter for AGS, GS and LISE, and 5 mm for LGSe, respectively. For sake of completeness the results of pulse compression discussed in section 4 are also added.

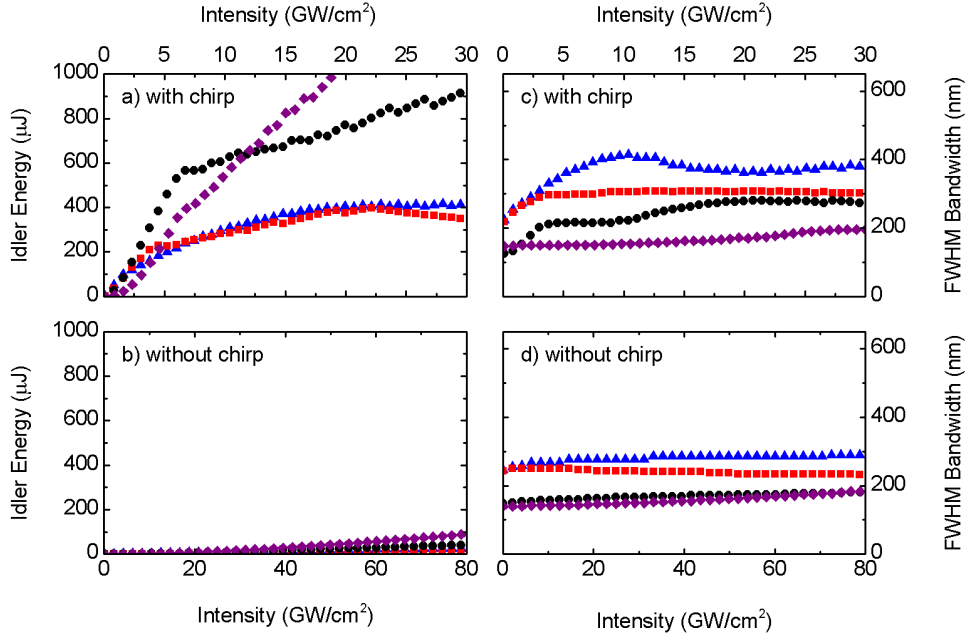

Fig. 2. Achievable output power and bandwidth for AGS (red squares), GS (blue triangles), LiSe (black dots), and LGSe (purple diamonds). (a) CP-DFG idler energy and (b) TL-DFG energy achieved with compressed pulses with the corresponding spectral bandwidth (c) and (d).

Table 1 3D simulation of maximum output energy and FWHM bandwidth for GS, AGS, and LISe and largest bandwidth for LGSe. The crystal dimensions are chosen to be 10 mm for AGS, GS and LISE and 5mm for LGSe, corresponding to the maximum commercially available crystal size. The parameters are optimized to get the maximum figure of merit described by equation 3.

| material | intensity (GW/cm$^2$) | chirp 760 nm / 850 nm (fs$^2$) | pump energy 760 nm / 850 nm (mJ) | output energy (µJ) | FWHM bandwidth (nm) | bulk Ge for compression (mm) | compressed pulse duration (fs) |
|---|---|---|---|---|---|---|---|
| 1 mm GS | 4.4 | 30750 / 22778 | 3.7 / 1.1 | 163 | 345 | 159 | 242 |
| 2 mm AGS | 2.2 | 70000 / 53846 | 3.5 / 1.7 | 107 | 281 | 438 | 286 |
| 3 mm LISe | 4.6 | 45000 / 38461 | 6.2 / 1.6 | 339 | 200 | 515 | 457 |
| 2 mm LGSe | 20.7 | 20000 / 15384 | 3.3 / 0.6 | 211 | 160 | 101 | 327 |

From Table 1 we can draw a few important conclusions. Under optimized mixing conditions (i.e. optimized crystal length, chirps and input energies, shown in the left half of the table) the two conventional nonlinear material AGS, and GS show the lowest output power, but the highest FWHM output bandwidth, while LISe and LGSe provide the largest output energy. With CP-DFG the output energy generated from AGS reaches up to 107 μJ, whereas in the un-chirped case only 6 μJ is produced. This value agrees well with experimental results from Xia et al. [15], where about 8 μJ were generated. Even though AGS and GS exhibit extremely strong TPA, we can achieve reasonable quantum efficiency with CP-DFG and high output energies of up to 163 μJ. Given the pump energy of 3.7 mJ, this corresponds to a quantum efficiency of 42%. In the un-chirped case one would expect a pulse of 21 μJ in the mid-IR (using optimized pump-signal energy of 3.2 mJ), corresponding to a quantum efficency of 14%. A similar improvement is observed for LISe and LGSe for which the quantum efficiency is improved from 14% to 50% and from 32% to 60%, respectively using CP-DFG. While AGS, GS, and LISe are available with 10 mm diameter, LGSe is currently only available with 5 mm diameter. Thus for the same TPA loss density significantly lower output energy can be achieved. The maximum input pulse energies for pump and signal are strongly varying for different crystals due to strongly varying TPA and different optimum chirps.

### 3. Optimization procedure

For a given nonlinear crystal the optimization of bandwidth and output energy is a multidimensional problem. Each possible input intensity requires a certain set of pump chirp, pump/signal chirp ratio, crystal length, and pump/signal intensity ratio. Since both input beams are relatively close in wavelength and divergence the two beams are chosen equal in size for sake of simplicity. The $1/e^2$ beam diameter was set equal to the crystal size to maximize throughput. Ideally one would need to run the optimization for intensities close to the damage threshold of the nonlinear material only. Unfortunately little is known about the corresponding damage threshold, thus our 1D calculations cover the intensity range, where we expect the damage limit (between 10 and 40 mJ/cm$^2$). For optimization we have applied the procedure as illustrated in Fig. 3, which we describe in the following.

In a first loop we optimize the pump/signal chirp ratio, then the nonlinear crystal length, and finally the pump/signal intensity ratio. The calculations are performed in 1D and in the plane wave approximation.

Since all parameter are linked to each other, the procedure consists of several optimization loops. In a first loop the optimum pump/signal chirp ratio is identified. For a selected pump/signal chirp ratio, we then calculate for each input intensity the required pump chirp for maximum conversion efficiency (ratio between generated idler energy and total pump and signal input energy). By this approach a large parameter space can be covered at reasonable calculation and evaluation time. Since the involved beams are relatively large with several millimeters of diameter, diffraction effects on the mixing process are neglected.

To restrict the parameter range and to prevent instabilities in the optimization procedure, the maximum pump chirp needs to be limited to an upper value. This limitation of maximum chirp is required since the pump chirp is inverse correlated to the total input intensity. While for increasing intensities the required chirp rate is decreasing and finally settling to a constant value with corresponding pulse duration between 1 and 2 ps (Fig. 4b.), the chirp rate at lower intensity is more problematic. For lower intensities the curves become indeed divergent (see e.g. for AGS below 5 GW/cm$^2$) and our optimization algorithm would simply pick the largest available pump chirp value since at this intensity levels the conversion efficiency becomes nearly independent of the chirp. The introduction of a maximum chirp rate prevents this instability. From an experimental point of view this limitation is rational since the maximum chirp is typically limited through the available optics in the grating compressor, for example.

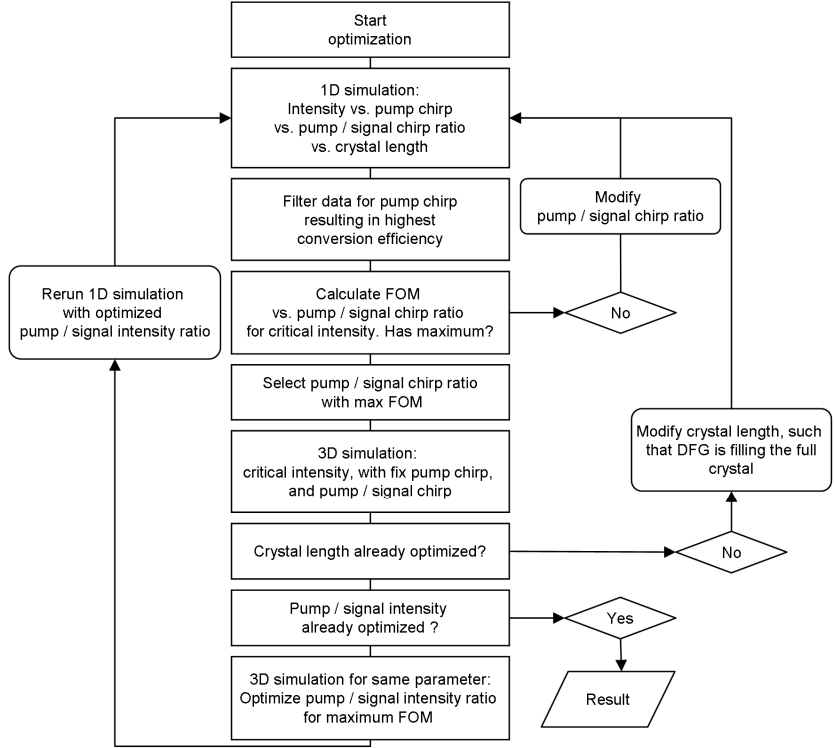

Fig. 3. Flowchart of multidimensional optimization procedure.

For a nonlinear mixing process one expects the largest FWHM output bandwidth for the following pump/signal chirp ratio $A_1$:

$$A_1 = 1 - \frac{n_{g,pump} - n_{g,signal}}{n_{g,idler} - n_{g,signal}} = \frac{GDD_{pump}}{GDD_{signal}}, \qquad (1)$$

where GDD is the group delay dispersion and $n_g$ the group velocity index of the interacting pulses. This relation is proposed by chirp assisted group velocity matching [9] which also suggests that the chirp ratio between pump and signal can be set to a fixed factor. The chirp ratio $A_2$ between signal and idler is then given by

$$A_2 = 1 - A_1 = \frac{GDD_{signal}}{GDD_{idler}}. \qquad (2)$$

This assures optimal phase matching bandwidth under the absence of TPA. In our case with stretched pulses, TPA is dependent of the spectral intensity and adds therefore an additional spectral loss component, thus competing with the gain through phase matching. There is no longer one chirp ratio that gives the best performance over the whole intensity range and the optimized value for $A_1$ needs thus been retrieved from our simulations. As selection criterion we define the following figure of merit (FOM)

$$FOM = \frac{E \cdot \Delta \lambda}{I}, \qquad (3)$$

where E is the generated idler pulse energy, $\Delta\lambda$ its FWHM spectral bandwidth and I the combined pump and signal intensity. The chirp ratio delivering the maximum FOM depends on the input intensity.

For further optimization the FOM needs to be evaluated at a fixed intensity. Since the damage threshold is unknown, we have defined a critical intensity. This selection is based on the following assumptions. We assume that energy dissipation through TPA is the dominating limiting effect. Since only little data on the subject is available we have based our definition on the earlier experimental implementation of TL-DFG performed by Xia et al. [13]. We estimated that in their experiment 75% of input energy was lost inside AGS due to TPA, corresponding to TPA induced loss density of 3.8 mJ/cm$^2$. We defined the critical intensity as the input intensity where the loss density is 3.8 mJ/cm$^2$ since Xia reported no serious damage on their crystal at this level. In general this value depends on the nonlinear crystal and serves thus here only as reference for crystal comparison.

By evaluating the FOM for the critical intensities the optimum pump/signal chirp ratio is found (Table 2). Compared to the value from chirp assisted group velocity matching, the optimum chirp factor is pushed to a higher value due to TPA. Pump and signal beam have the same sign in chirp, while the generated idler has the opposite chirp. In this article we have chosen positively chirped pump and signal in view of bulk-based pulse compression after CP-DFG.

Table 2 Parameter for optimum CP-DFG estimated from simulation

| Material | Crystal length (mm) | Optimum chirp factor | Chirp factor from GVM | Intensity ratio Signal / Pump | Figure of merit |
|---|---|---|---|---|---|
| GS | 1 | 1.35 | 1.25 | 0.48 | 12.8 |
| AGS | 2 | 1.30 | 1.25 | 0.82 | 13.7 |
| LISe | 3 | 1.17 | 1.17 | 0.38 | 14.7 |
| LGSe | 2 | 1.35 | 1.13 | 0.29 | 6.5 |

On the basis of the above found parameter sets for the critical intensity CP-DFG has been calculated in more details through 3D simulations and used for the second loop. This loop optimized the crystal length (Table 2), such that the conversion process is filling the whole nonlinear crystal, e.g. the energy of the idler reaches a maximum at the crystal end, no back conversion is taking place, and only a minor part of the pump energy is left. The result was then fed back into the first loop and reprocessed.

Initially one would expect the highest performance for equally intense pump and signal beams from the undepleted wave mixing approximation. CP-DFG is operating in the strongly depleted regime. Thus after crystal length optimization, the updated parameter set at critical intensity has been used to perform an optimization on the pump/signal intensity ratio for maximum FOM. The result was then fed back into the beginning of the procedure. The simulation was rerun with updated chirp ratio, crystal length, and pump/signal intensity ratio. The results are shown in Fig. 2. while the corresponding input pulse parameter are plotted in Fig. 4. and Fig. 5.

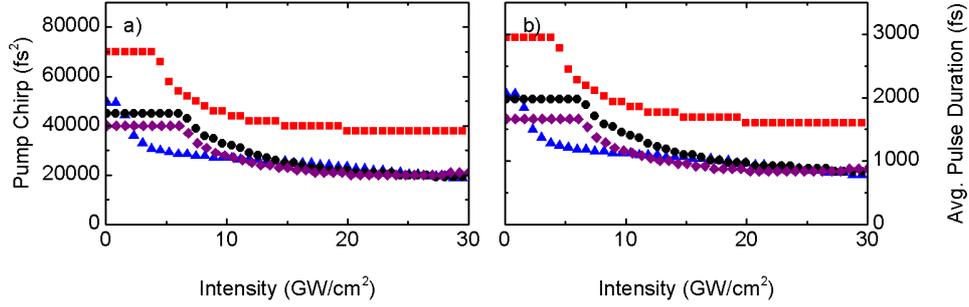

Fig. 4. (a) Required pump chirp value for maximum output energy at corresponding intensity. The chirp of the signal is set by the chirp factor from Table 2. (b) Corresponding average pulse duration of pump and signal pulses for AGS (red squares), GS (blue triangles), LISe (black dots) and LGSe (purple diamonds)

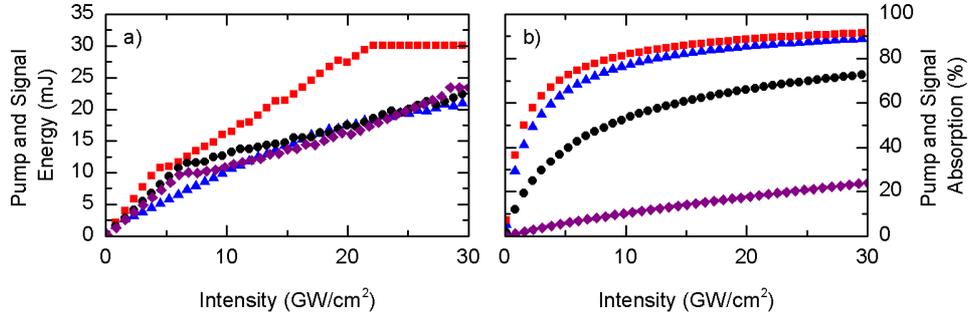

Fig. 5. (a) Total required pump energy at corresponding intensity for AGS (red squares), GS (blue triangles), LISe (black dots) and LGSe (purple diamonds). (b) Pump and signal absorption for CP-DFG due to two-photon absorption.

In principle, since pump and signal are relatively close in wavelength, we also expect a strong contribution from the cross TPA terms. Unfortunately little is known about these coefficients. To give an upper limit estimation we used therefore the same value as for the diagonal coefficients, corrected by a factor of 2 originating from the weak wave retardation [25]. To estimate the impact of nonlinear refractive index on our results, we need to rely on data available for other nonlinear materials. In literature only for a few materials both nonlinear refractive index and TPA coefficient are available at the same wavelength. Materials with a TPA coefficient between 1 and 10 cm/GW are e.g. silicon waveguides [26] or $AgGaSe_2$ (AGSe) [27], a close relative to AGS. Due to a lower band gap the TPA limit is shifted to longer wavelength and corresponding TPA coefficient is for 1.5 μm. The corresponding nonlinear refractive index are $6\times10^{-5}$ $cm^2$/GW for the silicon waveguides [26] and $3.5\times10^{-5}$ $cm^2$/GW for AGSe [28]. Thus we expect the nonlinear refractive index of GS and AGS for 800 nm in the same order of magnitude, and for LISE and LGSe correspondingly smaller. According to our simulations a nonlinear refractive index of $10\times10^{-5}$ $cm^2$/GW has no significant impact in GS, and AGS. We are therefore only expecting a minor impact on our result through a nonlinear refractive index since TPA dominates the conversion process by far. Other parasitic processes (e.g. thermo optic effect and linear absorption) have been neglected.

Experimentally the intensity needs to be set through the input energy for a given crystal size and pulse duration. Through the intensity ratio optimization more intensity is required in the pump than in the signal. Within the given intensity range, only the DFG output generated from AGS is limited by the available pump energy (Fig. 2a.). To access intensities above

20 GW/cm$^2$ tighter focusing needs to be applied. For the other nonlinear crystal sufficient input pulse energy is available to cover the investigated intensity range (Fig. 5.). The available laser energy from our Ti:sapphire system would allow higher intensities, but this would require larger nonlinear crystal.

## 4. Temporal pulse compression

The consequence of the CP-DFG approach is that the generated mid-IR pulse is strongly chirped. Recompression of mid-IR radiation is a challenging task. Due to the long wavelength relatively large beams are required for good collimation and alignment is challenging since the beam is difficult to visualize. Thus complicated prism or grating compressor configurations are time consuming to set up, require large components and in general add a significant loss. The large amount of dispersion needed to recompress a relatively narrow band mid-IR pulse (ca. 5% bandwidth) would require relatively large prism or grating separation, thus rendering it unfeasible. Therefore we present here recompression in bulk material. The amount of dispersion is estimated using equation (2). As mentioned earlier, the idler has negative chirp, while pump and signal have a positive chirp. Germanium has a positive GDD of 745 fs$^2$/mm at a wavelength of 7 µm. Thus relatively compact rods with a length of 200 mm can cover up to -150'000 fs$^2$ chirp, corresponding to a stretched pulse of up to 2 ps. The limitation of this compression scheme is that it does not allow for fine tuning. To achieve maximum compression, one needs to accurately adjust the chirp prior to compression. This can be done by adjusting the chirp of the two IR pulses, while maintaining the chirp ratio. This scheme does not allow controlling the third order dispersion. Due to the residual third order dispersion the generated output pulse has a small prepulse. In the case of GS, where potentially the shortest pulses can be generated, the bandwidth limited pulse duration is 232 fs (i.e. 10 cycles), while we can achieve 242 fs by bulk compression in 159 mm Germanium (Fig. 6b.).

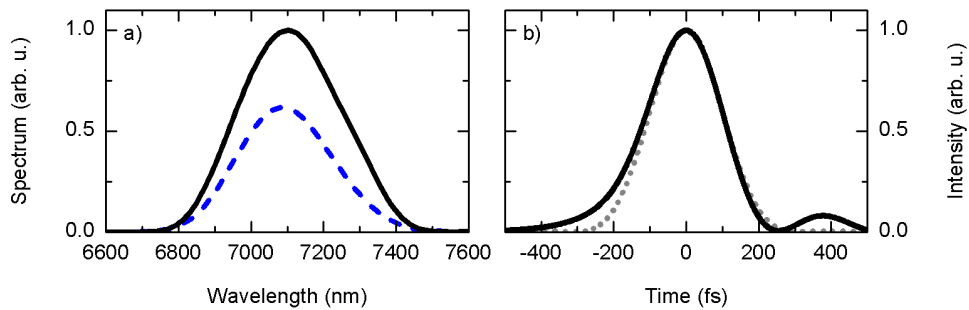

Fig. 6. Compressed idler output from CP-DFG in 1 mm GS with (a) spectral intensity at 0.4 mm inside the nonlinear crystal (blue dashed line) and at the output. (b) bandwidth limited (gray dotted line) and compressed (black solid line) pulse shape at the output after passing through 160 mm Germanium block.

The current design is based on positively chirped pump and signal. In principle the efficiency and bandwidth of the CP-DFG would be similar for negatively chirped pump and signal. The only difference would be that first the long wavelength components would interact and at the end the short wavelength parts. Pulse compression for this case seems more challenging since a material with negative dispersion in the mid-IR is required.

## 5. Conclusion

In conclusion we have designed a chirped pulse DFG stage that has the capability to efficiently mix the direct output from a dual wavelength Ti:sapphire amplifier system to the mid-IR. Compared to conventional un-chirped DFG we have shown by numerical simulation that the conversion efficiency can be significantly improved by a factor of 3 using CP-DFG. Based on different commercial available nonlinear crystals, the system is capable of delivering record-high 340 µJ pulse energy in 457 fs using LISe, and pulses as short as 242 fs (i.e. 10 cycles) with 165 µJ at 7.2 µm central wavelength. The two-color tunability offered by the Ti:sapphire system in combination with the investigated nonlinear material allows the emission of up to 16 µm. The high-energy pulses generated by CP-DFG are inherently CEP stable and are re-compressed in bulk Ge close to their transform-limited pulse duration. CP-DFG represents thus the potential for high-field applications with intense infrared pulses.

We would like to acknowledge fruitful discussions with Gunnar Arisholm, as well as the financial support from SNSF (Grant No. PP00P2_128493). CPH acknowledges association to NCCR-MUST, the National Centres of Competence in Research.